\documentclass[letters,fleqn,usenatbib,useAMS]{mnras}

\usepackage{graphicx}	

\newcommand{\logg}{\log\,g}
\newcommand{\teff}{T_{\rm eff}}
\newcommand{\bz}{\langle B_z \rangle}
\newcommand{\vsini}{$v\,\sin\,i$}

\title[Magnetic HgMn stars]{
Detection of weak magnetic fields in two HgMn stars
}

\stepcounter{footnote}
\author[Hubrig et al.]{
S.~Hubrig$^{1}$\thanks{Corresponding author: shubrig@aip.de},
S.~P.~J\"arvinen$^{1}$,
H.~Korhonen$^{2}$,
I.~Ilyin$^{1}$,
M.~Sch\"oller$^{3}$,
E.~Niemczura$^{4}$,
\and
S.~D.~Chojnowski$^{5}$
\\
$^{1}${Leibniz-Institut f\"ur Astrophysik Potsdam (AIP), An der Sternwarte~16, 14482~Potsdam, Germany} \\
$^{2}${European Southern Observatory, Alonso de Cordova 3107, Vitacura, Santiago, Chile} \\
$^{3}${European Southern Observatory, Karl-Schwarzschild-Str.~2, 85748 Garching, Germany} \\
$^{4}${Astronomical Institute, University of Wroc{\l}aw, Kopernika 11, 51-622 Wroc{\l}aw, Poland} \\
$^{5}${Apache Point Observatory and New Mexico State University, PO Box 59, Sunspot, NM 88340-0059, USA} \\
}

\date{Accepted XXX. Received YYY; in original form ZZZ}

\pubyear{2019}

\begin{document}
\label{firstpage}
\pagerange{\pageref{firstpage}--\pageref{lastpage}}
\maketitle

\begin{abstract}
The main-sequence mercury-manganese (HgMn) stars are known to exhibit large overabundances
of exotic elements and, similar to magnetic Ap/Bp stars,
are spectrum variables, implying the presence of an inhomogeneous element distribution over the stellar
surface. A number of magnetic field studies have been attempted in the last decades,
indicating that
magnetic fields in HgMn stars, if they exist, should be rather weak.
The presence of tangled magnetic fields was suggested by several authors who detected
quadratic magnetic fields using the moment technique.
We employ the least-squares deconvolution technique 
to carry out a sensitive search for weak magnetic fields in
spectropolarimetric observations of three HgMn stars, HD\,221507, HD\,65949, and HD\,101189, which have 
different fundamental parameters and spectral characteristics.
A definite weak longitudinal field
is discovered in HD\,221507 and HD\,65949 on single epochs, while only marginal field detections
were achieved for HD\,101189.
The new measurements
indicate that the structure of the magnetic fields is probably rather complex: our analysis
reveals the presence of reversed Stokes~$V$ profiles at the same observational epoch
if individual elements are used in the measurements.
This is the first observational evidence that individual elements
sample distinct local magnetic fields of different polarity across the stellar surface.
\end{abstract}


\section{Introduction}
\label{sec:intro}

The main-sequence B-type stars with spectral classes B7--B9 comprise two groups of 
chemically peculiar stars, the classical magnetic peculiar Bp stars and the HgMn stars. 
The group of Bp stars is composed of Si and He-weak stars,
which are usually characterised by large overabundances of Fe-peak and rare-earth 
elements. They possess non-axisymmetric, large-scale magnetic fields of up
to several kG and display on their surface chemical spots of different elements.
The most distinctive features of the HgMn group are the extreme overabundances of heavy elements,
in particular of Hg, which can reach up to 5-6\,dex, and/or Mn, up to 3\,dex. 
Other overabundance anomalies 
were detected for the chemical elements Y, P, Ga, Cu, Be, Bi, Zr, Sr, Xe, Au, and Pt,
whereas the elements 
He, Al, Ni, and Co are typically underabundant \citep[e.g.,][]{Smith1993,Castelli2004a,Hubrig2011}.  
Notably, as of today, no theory 
can account satisfactorily for the observed abundance pattern in HgMn stars. It remains still difficult to
 reproduce in numerical models the detailed abundances of these stars, although some progress was achieved
 in the understanding of the dependence of Mn overabundances on $\teff$ (e.g.\ \citealt{alecian1981,alecian2019}).

Using a survey of HgMn stars in close spectroscopic binaries, \citet{Hubrig1995}
were the first to discuss the existence of an inhomogeneous distribution
of some chemical elements over the surface of HgMn stars,
in particular with a preferential concentration of Hg along the equator.
Recent studies  have  revealed that not  only Hg,  but also  many  other  elements,  most
typically Ti, Cr, Fe, Mn, Sr, Y, and Pt, are concentrated in spots of diverse sizes,
and that different elements exhibit different abundance distributions across the
stellar surface \citep[e.g.,][]{Hubrig2006a,Briquet2010,Korhonen2013}.
Moreover, a few detailed studies based on extensive spectroscopic material reported
the  presence  of dynamical  abundance spot evolution
of  several  elements  at  different  timescales \citep[e.g.,][]{Briquet2010,Korhonen2013,Hubrig2010}.

The question of the presence of weak magnetic fields in HgMn stars is still under debate.
It was suggested that the difficulty of spectropolarimetric studies to unambiguously
detect the fields in HgMn stars can be attributed to the complexity of the surface magnetic 
field topology \citep{Hubrig1998a}.
Highly structured magnetic field topologies can reduce the Stokes~$V$
signatures in  spectral  line profiles below the detection  threshold. 
Furthermore, since HgMn stars usually display element spots on 
their surface, it is advisable to use in the magnetic field analysis 
line masks for individual elements separately. Combining spectral
lines of various elements may lead to the dilution of the magnetic signal or even to its (partial) 
cancellation, if enhancements of different elements occur in regions of opposite magnetic polarity. 

Attempts to detect mean
longitudinal magnetic fields in HgMn stars using low- and high-resolution spectropolarimetry have  been  made
by several authors, who reached contradictory results. While low-resolution observations
do not allow the use of individual elements separately, the analysis of polarimetric spectra 
using the moment technique \citep{Mathys1993} applied to individual elements was indicative of the presence 
of weak longitudinal magnetic field crossover and quadratic fields  \citep[e.g.,][]{Mathys1995,Hubrig1998a,Hubrig2012}.
To achieve higher accuracies in the measurements of weak magnetic fields,
a line addition technique, 
the least-squares deconvolution \citep[LSD;][]{Donati1997}, combining hundreds and sometimes up to thousand
lines of all elements is frequently applied. The main assumption in LSD is the application of the weak 
field approximation, that is,
the magnetic splitting of spectral lines is assumed to be smaller than their Doppler
broadening. Furthermore, it is assumed that the local line profiles are self-similar
and can be combined into an average profile. Due to non-linear effects in the 
summation and the effect of blends, the resulting LSD profiles should not be considered
as observed, but rather processed Zeeman signatures. 
Using the lines of different elements together,
\citet{Makaganiuk2011} were not able to obtain any definite detection of a mean longitudinal magnetic field 
in a sample of HgMn stars,
but could decrease statistical uncertainties down to below 10\,G for the best cases.
However, combining the lines of elements with different surface abundance distributions unavoidably leads to  
the dilution of  the magnetic signal, which is expected to be rather weak in HgMn stars.
Just in one case, the LSD study of the HgMn star HD\,11753, did the authors use
line lists for Ti, Y, and Cr separately \citep{Makaganiuk2012}. These measurements indicated a mean longitudinal 
magnetic field in the range from about $-20$\,G to $+25$\,G, with measurement uncertainties of up to 25\,G. 

In the following sections, we present the LSD analysis of spectropolarimetric observations of three HgMn stars 
with different fundamental parameters 
and spectral characteristics, HD\,221507 (=$\beta$\,Scl), HD\,65949, and HD\,101189 (=HR\,4487).
HD\,221507 is of special interest as a flare was discovered in the course of the inspection of 
Transiting Exoplanet Survey Satellite ({\em TESS}) light curves by \citet{Balona2019}.
According to \citet{arcos2017}, HD\,221507 is not a Be star,
as it lacks H$\alpha$ emission and has a low \vsini.
The fundamental parameters $\teff=12\,400$\,K and $\logg=3.90$ were determined by \citet{Alecian2016}.
Furthermore, \citet{Balona2019} detected the rotation frequency for this star, $\nu_{\rm rot}=0.522$\,c/d
using high-precision short-cadence (2\,min) photometric observations assembled by TESS \citep{Balona2019}.
In spite of the fact that spectral variability was already detected in numerous spectroscopical studies 
of HgMn stars, only very few HgMn stars have well determined rotation periods. 
HD\,221507 was observed using the High Accuracy Radial velocity 
Planet Searcher polarimeter \citep[HARPS\-pol;][]{snik2008} on ten different epochs from 2010 to 2019.

\begin{table}
\centering
\caption{
Observation log of the three HgMn stars.
 The columns list the heliocentric Julian date (HJD) for
  the middle of the exposures and the signal-to-noise ratio ($S/N$) of the
  spectra measured at about 5200\,\AA{}.
}
\label{tab:obs1}
\begin{tabular}{r r r r}
\noalign{\smallskip}
\hline
\hline
\noalign{\smallskip}
\multicolumn{1}{c}{HJD} &
\multicolumn{1}{c}{$S/N$} &
\multicolumn{1}{c}{HJD} &
\multicolumn{1}{c}{$S/N$} \\
\multicolumn{1}{c}{2\,400\,000+} &
&
\multicolumn{1}{c}{2\,400\,000+} &
\\
\noalign{\smallskip}
\hline
\noalign{\smallskip}
\multicolumn{2}{c}{HD\,221507} &
	\multicolumn{2}{c}{HD\,65949} \\
55210.549  & 526 &
	55600.708 & 166 \\
56128.935  & 541 &
	58649.479 & 147 \\
56146.791  & 186 &
	58652.477 & 86 \\
56491.678  & 464 &
	\\
56491.898  & 581 &
	\multicolumn{2}{c}{HD\,101189} \\
56492.667  & 578 &
	54983.474 & 63 \\
56492.896  & 505 &
	55201.863 & 456 \\
58646.936  & 365 &
	56491.521 & 581 \\
58650.810  & 370 &
	56492.481 & 599 \\
58652.931  & 406 &
	\\
\noalign{\smallskip}\hline \noalign{\smallskip}
\end{tabular}
\end{table}

\begin{figure*}
 \centering 
\includegraphics[width=0.70\textwidth]{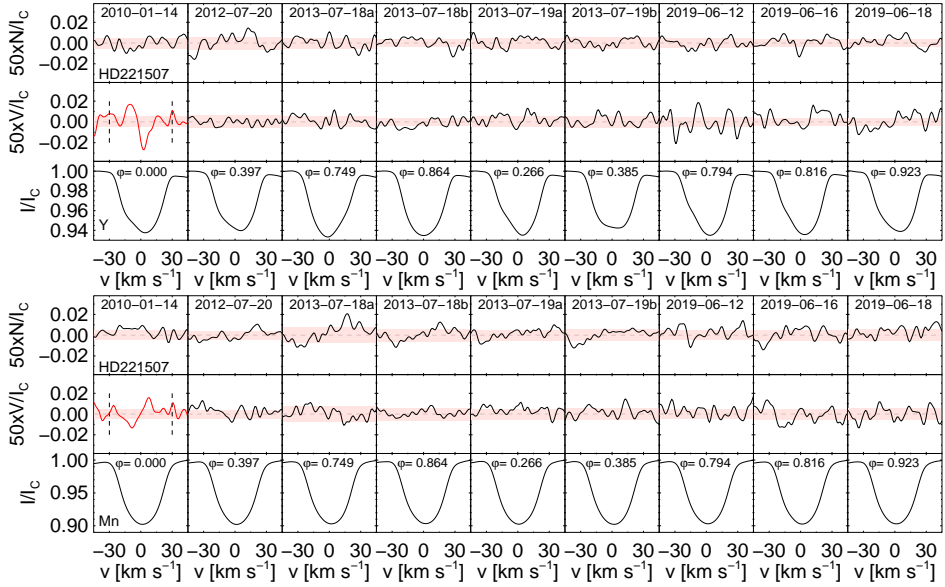}        
        \caption{
          LSD Stokes~$I$ (bottom), Stokes~$V$ (middle), and diagnostic null (N)
          profiles (top) for HD\,221507 using the Y line mask (upper panel) and the Mn line mask (bottom panel).
          The widths of the shaded red bands indicate the $\pm1\sigma$ ranges. Corresponding rotation phases for
          each observation are presented on top of the Stokes~$I$ profiles.
          }
   \label{fig:mos2}
\end{figure*}

\begin{figure}
 \centering 
\includegraphics[width=0.48\textwidth]{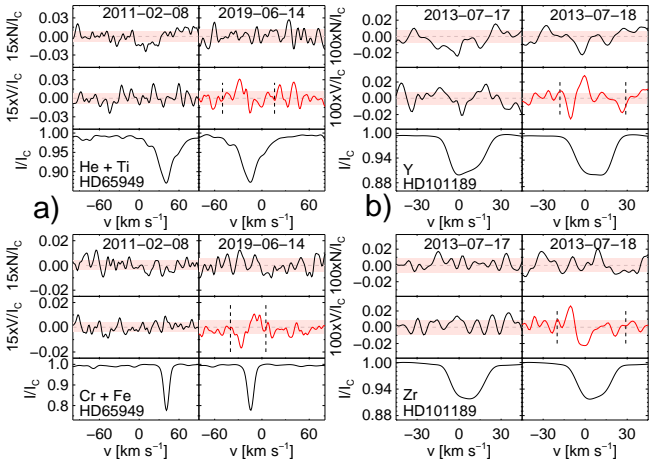}        
        \caption{
LSD Stokes~$I$ (bottom), Stokes~$V$ (middle), and diagnostic null (N) profiles (top).
         {\it Left side:}
Results for HD\,65949 using the He+Ti line mask (upper panel) and the Cr+Fe line mask (lower panel).
{\it Right side:}
Results for HD\,101189 using the Y line mask (upper panel) and the Zr line mask (lower panel).
}
   \label{fig:mos3}
\end{figure}

The star HD\,65949 with the atmospheric parameters  $\teff=13\,100$\,K and $\logg=4.0$ \citep{Alecian2016} 
was studied in detail by \citet{Cowley2010}. It belongs to the young open
cluster NGC\,2516 and is a SB1 system with an orbital period of 21.2\,d. Interestingly, the authors
also detected a small variation in the centre-of-mass velocity, which was
interpreted as due to the presence of a third body. This makes HD\,65949 a triple system. 
HD\,65949 exhibits enormous enhancements of the elements rhenium and osmium
through mercury (${\rm Z}=75-80$). According to \citet{Cowley2010},  osmium  and  rhenium are rarely enhanced to the
point where they are identified in ground-based spectra of HgMn stars. The authors suggested that
the high abundance in HD\,65949 must result from a very recent transfer of material. 
HD\,65949 was observed with HARPS\-pol on three different epochs from 2011 to 2019. 

HD\,101189 with the atmospheric parameters $\teff=11\,020$\,K and $\logg=3.9$ \citep{Alecian2016} is not known to belong 
to a SB system, but seems to be a member of a visual binary. \citet{Scholler2010} detected a close companion candidate
at a separation  of  0.337\arcsec{} and a position  angle of $104.1\degr$ using adaptive optics near-infrared observations.
HD\,101189 appears to be a good representative of typical HgMn stars, exhibiting a distinct variability of 
line profiles of several elements \citep{Hubrig2011}. 
HD\,101189 was observed with HARPS\-pol on four different epochs from 2009 to 2013.

\section{Data reduction and magnetic field measurement results}
\label{sect:obs}

HARPS\-pol observations of the three HgMn stars were
obtained  during  several observing campaigns organized by different groups of observers from 2009 to 2019. 
Each spectropolarimetric observation usually consists of
subexposures observed at different positions of the quarter-wave retarder plate. 
The resolving power of HARPS is about
$115\,000$ and the spectra cover the spectral range 3780--6910\,\AA{},
with a small gap between 5259\,\AA{} and 5337\,\AA{}. The data reduction 
was performed using the HARPS\-pol data reduction
software available on La~Silla and at ESO Garching. The normalization of the spectra to the
continuum level is described in detail by 
\citet{Hubrig2013}.
The heliocentric Julian date (HJD) for the middle of the exposures and the corresponding signal-to-noise ratio 
($S/N$) of the HARPS Stokes~$I$ spectra measured at about 5200\,\AA{} are presented in Table~\ref{tab:obs1}.

\begin{figure}
 \centering 
        \includegraphics[width=0.38\textwidth]{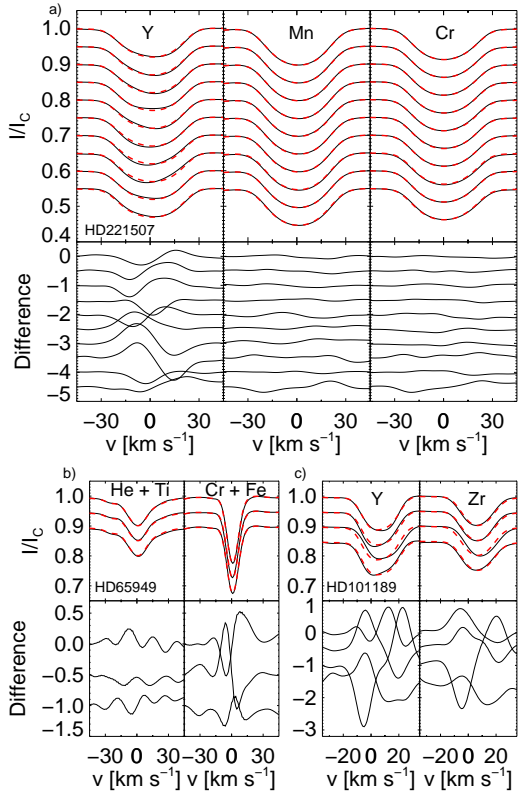}
        \caption{
Variability of the LSD Stokes~$I$ profiles belonging to different elements in HD\,221507 (a), HD\,65949 (b), and 
HD\,101189 (c).
The profiles obtained on different nights are overplotted by the mean profiles (red dashed lines) and 
offset vertically for better visibility. They are presented in chronological order with the oldest
observations at the bottom.
The lower panels present the differences between the individual SVD Stokes~$I$ profiles and the average 
profiles, multiplied by a factor of 100.
}
   \label{fig:mos1}
\end{figure}

Line masks for a number of elements known from several previous studies to be inhomogeneously distributed 
over the surface of HgMn stars were obtained from the
Vienna Atomic Line Database \citep[VALD; e.g.,][]{VALD3}. They were constructed for each star 
based on their stellar parameters given in Sect.~\ref{sec:intro}.
The considered line masks  belong to the elements He, Ti, Cr, Fe, Mn, Y, and  Zr, 
which frequently show an inhomogeneous distribution on the surface of HgMn stars.
The major problem in the search for very weak magnetic fields
is the proper line identification of blend free spectral lines. The
quality of the selection varies strongly from star to star, depending
on binarity, line broadening, and the richness of the spectrum.
We have checked that each line in the selected line mask is present in the stellar spectra.
Obvious line blends and lines in telluric regions were excluded from the line
masks.
The mean longitudinal magnetic field is calculated by computing the
first-order moment of the Stokes~$V$ profile according to \citet{Mathys1989}.

The resulting LSD profiles for the HARPS\-pol spectra obtained with the best $S/N$  are shown in Figs.~\ref{fig:mos2} and 
\ref{fig:mos3}.
To classify the magnetic field detection in the LSD technique, False Alarm Probabilities 
(FAPs) are commonly used \citep{Donati1992}: a profile 
with ${\rm FAP} < 10^{-5}$ is considered a definite detection, with $10^{-5} < {\rm FAP} < 10^{-3}$ a 
marginal detection, and with ${\rm FAP} > 10^{-3}$ a non-detection.
A positive mean longitudinal magnetic field $\bz=35\pm4$\,G (${\rm FAP} < 2.8\times10^{-7}$), is definitely 
detected in the spectra of HD\,221507 acquired in 2010 (HJD\,2455210.549) using the Y line mask.
Only a marginal detection of a field of negative polarity, 
$\bz=-16\pm3$\,G (${\rm FAP} < 1.6\times10^{-5}$) was achieved at the same epoch using the Mn line mask.
Interestingly, \citet{Hubrig2012} applied the moment technique to a sample of Y lines observed in the HARPS\-pol
spectra obtained at the same epoch and reported
$\bz=78\pm25$\,G, which is in good agreement with the presented LSD measurements for this element,
$\bz=35\pm5$\,G, within the error bars.  
The variability of the spectral lines of a few elements is presented in Fig.~\ref{fig:mos1}a.
Obviously, the most variable lines in the spectra of this star belong to Y.
 
HD\,65949 exhibits the strongest longitudinal 
magnetic field up to $\bz=187\pm36$\,G (${\rm FAP} < 7.1\times10^{-6}$) 
measured using the He$+$Ti line mask, and up to $\bz=-130\pm22$\,G (${\rm FAP} < 2.3\times10^{-7}$)
using the Cr$+$Fe line mask. The detection of a mean longitudinal magnetic field 
in this star is in agreement with the previous detection ($\bz=-290\pm62$\,G) by \citet{Hubrig2006b} who used
the FOcal Reducer low dispersion Spectrograph \citep[FORS\,2;][]{Appenzeller1998},
installed at the ESO/VLT, in spectropolarimetric mode.
The variability of the lines in both masks is displayed in Fig.~\ref{fig:mos1}b.

HD\,101189 is Y and Zr rich and shows pronounced spectral variability (Fig.~\ref{fig:mos1}c).
According to \citet{Hubrig2010}, also many other elements show variability of their line profiles.
The analysis yields marginal detections with
$\bz=7\pm4$\,G (${\rm FAP} < 7.1\times10^{-5}$) using the Y line mask and 
$\bz=9\pm5$\,G (${\rm FAP} < 1.2\times10^{-4}$) using the Zr line mask.

No indication of a Zeeman signature was found
in any null spectrum, yielding formal non-detections with ${\rm FAP} > 10^{-3}$.
Null spectra are usually obtained by combining the 
different polarizations in such a way that polarization cancels out.

\section{Discussion}
\label{sec:meas}

The detection of a positive polarity of the longitudinal magnetic field $\bz$ measured using the Y line mask and
the marginal detection of a field of negative polarity measured using the Mn line mask in the first HARPS\-pol 
observation of HD\,221507 acquired in 2010 (HJD\,2455210.549) indicates the possible presence of local tangled
magnetic fields, which are sampled differently by different elements.
Similar to the results of our analysis of HD\,221507, opposite field polarities are detected using different
line masks for HD\,65949. 
A number of studies using the moment technique already indicated the presence
of tangled magnetic fields in a few HgMn stars.
However, due to the weakness of the longitudinal magnetic fields in these stars,
the detections of the presence of complex fields was exclusively based on the discovery of 
quadratic magnetic fields \citep[e.g.,][]{Mathys1995,Hubrig1998a,Hubrig2012}. 
The presence of opposite field polarities in temperature spots in the late-type star II\,Peg
was recently discussed by \citet{stras2019} and appears to be typical in magnetically active
late-type stars.
However, the detection of opposite field polarities 
using line masks for different elements is unexpected. Although it is tempting to assume that the 
detection of a flare in HD\,221507 by \citet{Balona2019} 
is related to the presence of a complex magnetic field structure similar to that typically observed in active 
low-mass stars, there is a possibility that the observed flare originated in the close-by late-type companion
at a separation of 0.641\arcsec{} detected by \citet{Scholler2010}.
On the other hand, previous observations of 13 HgMn stars,
including HD\,221507,
with the ROSAT High-Resolution Imager confirmed X-ray sources at the position of these stars \citep{Hubrig1998b}.
Future high spatial
resolution observations (e.g.\ with {\sl Chandra}) are needed to determine whether the origin of the observed X-ray emission
is related to the HgMn stars themselves or to late-type companions.

Since we have not discovered magnetic fields in all the observations
obtained at different epochs distributed over several years,
it is quite possible that their detectability strongly depends on the spot position and the element abundance 
within the spot, i.e.\ we probably observe different ``evolutionary states'' of the spots, especially 
when observing epochs are separated by long time intervals.

In our analysis we detect that Zeeman features in the  Stokes~$V$ profiles of HD\,65949 and HD\,101189
appear shifted bluewards.
This probably indicates that the observed polarization
signal corresponds to a fraction of the visible hemisphere and could be related to a surface 
inhomogeneous distribution of the individual chemical elements. Similar blueward shifts of Zeeman features
persisting for more than half a rotation period were also observed in the magnetic Bp star 36\,Lyn showing
surface element patches \citep{wade2006}.
If the measured magnetic fields are associated with 
an inhomogeneous distribution of the elements on the stellar surface,
then the field strengths measured in the longitudinal strips present just a lower limit for the field strength 
expected to be measured in the chemical spots directly.

The possibility that the detected Zeeman features originate
in close-by late-type companions to these stars can be discarded as
the visual close companion to HD\,101189 was never detected spectroscopically.
According to the orbital radial velocities of HD\,65949 presented by \citet{Cowley2010},
the observed Zeeman feature cannot be related to the secondary of the SB1 system. The spectrum can still
be contaminated by the distant third component, but apart from the center-of-mass velocity changes there is no
trace of this component in the available spectra.

Possible scenarios for the generation of magnetic fields in HgMn stars, which are frequently in binaries and
multiple systems, were already discussed in the past by \citet{Hubrig2011}.
While classical magnetic Bp stars in close 
binaries are very rare, more than 2/3 of the HgMn stars are known
to belong to spectroscopic binaries, with a
preference of orbital periods ranging from 3 to 20 days \citep{Hubrig1995}.
Surveying a magnitude limited sample of late B-type stars
using the ninth catalogue of spectroscopic binary 
orbits by \citet{pourbaix2004}
reveals only very few normal late-B type primaries in systems with orbital periods below 20\,d
\citep{Hubrig1995}.
Based on these statistics, it cannot be excluded that most late B-type stars formed in binary
systems with certain orbital parameters become HgMn stars \citep{Hubrig2011}.
\citet{Hubrig2011} 
suggested that a tidal torque varying with depth and latitude in a star induces 
differential rotation. Differential rotation in radiative chemically peculiar stars 
can be prone to magneto-rotational
instability (MRI). Magnetohydrodynamical simulations by \citet{Arlt2003} revealed a distinct structure for the
magnetic field topology similar to the fractured  elemental rings observed on the surface of HgMn stars.
The introduction of a magnetic field excites the MRI on a very short time-scale when compared to
the time-scale of microscopic magnetic diffusion.
Although the fields are not strong, complex surface patterns can be obtained from the nonlinear,
nonaxisymmetric evolution of the MRI.

Due to the weakness of the detected magnetic fields, future observing campaigns should be based on 
high-resolution spectropolarimetric observations obtained at a very high $S/N$
and have to be carried out during rather short time intervals of the order of a few weeks:
As has already been shown by \citet{Briquet2010} and \citet{Korhonen2013}, 
chemical spots on the surface of HgMn stars are not stable and show temporal evolution of the spot shape and 
the element abundance on time scales as short as a few months.  \citet{Briquet2010} found that Y, Sr, and Ti
spots  change  their  configuration  on  HD\,11753  between two data sets that
were separated by approximately two months, thus providing clear evidence of
dynamical spot evolution on a short timescale.
\citet{Korhonen2013} investigated further the reliability of that fast dynamical
spot evolution reported for HD\,11753, and confirmed, both from chemical spot
maps and from equivalent width measurements, that the spot configuration
changed on a timescale of months.

\section*{Acknowledgements}
We would like to dedicate this work to our late colleague Dr. Fiorella Castelli.
We thank the referee for the insightful questions and
suggestions.
Based on observations made with ESO Telescopes at the La Silla Paranal
Observatory under programme IDs 083.D-1000, 084.D-0338, 086.D-0240, 187.D-0917, 089.D-0383, 091.D-0759, 
and 0103.C-0240.
This work has made use of the VALD database, operated at Uppsala University,
the Institute of Astronomy RAS in Moscow, and the University of Vienna,
and of the SIMBAD database, operated at CDS, Strasbourg, France.

\bsp	
\label{lastpage}
\end{document}